\documentstyle[preprint,pra,aps,eqsecnum,epsf]{revtex}

\begin{document}
\preprint{                                                 CERN-TH/99-286}
\tightenlines

\title{Solar Mikheyev-Smirnov-Wolfenstein Effect\\
 with Three Generations of Neutrinos}

\author{Per Osland}

\address{Department of Physics, University of Bergen,
All\'{e}gaten 55, N-5007 Bergen, Norway}

\author{and}
\author{Tai Tsun Wu}

\address{Gordon McKay Laboratory, Harvard University, Cambridge,
Massachusetts 02138, and\\
Theoretical Physics Division, CERN, CH-1211 Geneva 23, Switzerland}

\maketitle

\begin{abstract}
\indent Under the assumption that the density variation of the electrons
can be approximated by an exponential function, the solar
Mikheyev-Smirnov-Wolfenstein effect is treated for three
generations  of neutrinos. The generalized hypergeometric
functions that result from the exact solution of this problem are
studied in detail, and a method for their numerical evaluation is
presented. This analysis plays a central role in the determination
of neutrino masses, not only the differences of their squares,
under the assumption of universal quark-lepton mixing.
\end{abstract}

\vfill
\begin{flushleft}
 {CERN-TH/99-286} \\[2mm]
 {December 1999}
\end{flushleft}
\eject
\setcounter{page}{1}

\section{Introduction}

Recently a program has been proposed to study neutrino masses
\cite{OW-99-L}.
This program is based on the following assumptions.

\noindent {\bf Assumption 1:} There are three right-handed Dirac
neutrinos in addition to the three known left-handed neutrinos.

\noindent {\bf Assumption 2:} The lepton mass matrices have the
same structure as the quark mass matrices, as described by Lehmann
{\it et al.} \cite{LNW}.

In carrying out this program, one of the technically most difficult
aspects is to treat the Mikheyev-Smirnov-Wolfenstein (MSW) effect
for the sun \cite{MSW}.  It is the purpose of this paper to discuss
this problem.

The difficulty arises in the following way. On the basis of
the Super-Kamiokande data on atmospheric neutrinos \cite{Super-K-prl},
the muon neutrino $\nu_\mu$ oscillates with another neutrino which is
not the electron neutrino $\nu_e$.  From Assumption~1, this neutrino must be
the tau neutrino $\nu_\tau$.  In other words, $\nu_\mu$ and $\nu_\tau$ are
coupled significantly.  Again from Assumption~1, the electron neutrinos
$\nu_e$ from the sun must therefore oscillate with either $\nu_\mu$ or
$\nu_\tau$.  Since $\nu_\mu$ and $\nu_\tau$ are coupled significantly, the
electron neutrinos $\nu_e$ from the sun must oscillate with {\it both}
$\nu_\mu$ {\it and} $\nu_\tau$. It is accordingly necessary to understand
the MSW effect with all three generations of neutrinos $\nu_e$, $\nu_\mu$
and $\nu_\tau$.

While there is an extensive literature on the MSW effect for two
generations of neutrinos \cite{MSW-2-exp,MSW-2-lin}, relatively little is
known for the case of three generations \cite{MSW-3}.  
One of the major differences between these two cases is the following.  
For two generations, the MSW effect occurs predominantly in the vicinity of 
a particular density in the sun.  For this reason, the important parameter
is the rate of change of the density at this particular density.  In
contrast, for three generations, the important contributions to
the MSW effect can come from several regions.  When the parameters, such
as the neutrino masses, vary, these regions not only move but also merge
and separate, features that are absent for the much simpler case of two
generations.

\section{Formulation of the Problem}

There are many different ways to formulate the problem of the MSW
\cite{MSW} effect in the sun for three generations of neutrinos.  These
different formulations are of course closely related to each other; it has
been found that, at least for the present program \cite{OW-99-L}, the
following one is most appropriate.

In a current basis, the charged lepton currents are
\begin{equation}
J_\mu\sim \bar\nu_{\rm L}\gamma_\mu \ell_{\rm L},
\end{equation}
where
\begin{displaymath}
\nu=
\left[\tighten\matrix{\nu_1 \cr \nu_2 \cr \nu_3\cr
}\right] \quad
{\rm and} \quad
\ell=\left[\tighten\matrix{
\ell_1 \cr \ell_2 \cr \ell_3\cr}\right]
\end{displaymath}
denote the neutrino and charged lepton fields.  The neutrino mass matrix
$M$ \cite{Rosen}, 
from the vacuum expectation value of the Higgs field in the
neutrino-Higgs coupling, is
\begin{equation}
\bar\nu_{\rm L} M \nu_{\rm R}+{\rm h.c.}
\end{equation}

Let $p$ be the momentum of the neutrino, the value of which is taken to be
much larger than the neutrino masses, which are the eigenvalues of $M$.
Typically, $p$ is in the range from 0.4~MeV/$c$ to 10~MeV/$c$,
while the neutrino masses are believed to be no more than about 1~eV;
hence this assumption is well satisfied.
Under this assumption, it is $M^2$ that enters in the MSW effect.
Let
\begin{equation}
M^2
=\left[\tighten\matrix{
M^2_{11} & M^2_{12} & M^2_{13}\cr
M^2_{21} & M^2_{22} & M^2_{23}\cr
M^2_{31} & M^2_{32} & M^2_{33}\cr}\right]
\end{equation}
be real and symmetric.
Note that $M_{ij}$ are not defined, only $M^2_{ij}\equiv(M^2)_{ij}$.
Of course the eigenvalues of this matrix $M^2$ are the squares
of the neutrino masses.

In terms of this $M^2$, the MSW effect for neutrino oscillations is
described by the coupled ordinary differential equation \cite{MSW}
\begin{equation}
\label{Eq:Schr-1}
i\,\frac{d}{dr}
\left[\tighten\matrix{
\phi_1(r)\cr \phi_2(r)\cr \phi_3(r)\cr}\right]
=\left(
\left[\tighten\matrix{
D(r) & 0 & 0\cr
0 & 0 & 0\cr
0 & 0 & 0\cr}\right]
+\frac{1}{2p}
\left[\tighten\matrix{
M^2_{11} & M^2_{12} & M^2_{13}\cr
M^2_{21} & M^2_{22} & M^2_{23}\cr
M^2_{31} & M^2_{32} & M^2_{33}
\cr}\right]
\right)
\left[\tighten\matrix{
\phi_1(r)\cr \phi_2(r)\cr \phi_3(r)
\cr}\right],
\end{equation}
where
\begin{equation}
\label{Eq:define-D}
D(r)=\sqrt{2}\,G_{\rm F} N_e(r),
\end{equation}
with $G_{\rm F}$ the Fermi weak-interaction constant
and $N_e(r)$ the electron density at the point $r$.

So far this formulation is general.
In view of the observation in Sec.~I that, for three generations
of neutrinos, important contributions to the MSW effect can come from
several values of $r$, we propose to choose a suitable function $D(r)$.
On the one hand, this $D(r)$ should describe approximately the electron
density in the sun; on the other hand, this $D(r)$ should be sufficiently
simple so that Eq.~(\ref{Eq:Schr-1}) can be solved explicitly.
This choice is made in the following way.

First, since most neutrinos are produced by nuclear reactions \cite{sun}
near the center of the sun, the simplifying assumption is made that the
neutrinos are produced at the center of the sun.
This has the consequence that the $r$ in Eqs.~(\ref{Eq:Schr-1})
and (\ref{Eq:define-D}) is the radial distance from the center of the sun.
This is the reason for the notation $r$.

Secondly, the density of the sun as a function of the radial distance
is fairly accurately known \cite{BBP-98}, as shown by the solid curve
in Fig.~1, where the vertical axis is the density in g/cm$^3$ in a
logarithmic scale, and the horizontal axis is $r/R_\odot$.
The radius of the sun is
\begin{displaymath}
R_\odot=6.96\times10^8\ {\rm m}.
\end{displaymath}
 From this solid curve in Fig.~1, it is seen that the solar density
is approximately exponential, as shown by the dotted line of Fig.~1.
This fit is reasonably good, to an accuracy of roughly 15\%,
for
\begin{displaymath}
0.05 < r/R_\odot < 0.9.
\end{displaymath}
In other words, $D(r)$ is taken to be given by
\begin{equation}
\label{Eq:D-exponential}
D(r)=D(0)e^{-r/r_0}.
\end{equation}
\begin{figure}[bh]
\refstepcounter{figure}
\label{Fig:rho}
\addtocounter{figure}{-1}
\begin{center}
\setlength{\unitlength}{1cm}
\begin{picture}(12,7.5)
\put(2.0,1.0)
{\mbox{\epsfysize=7.cm\epsffile{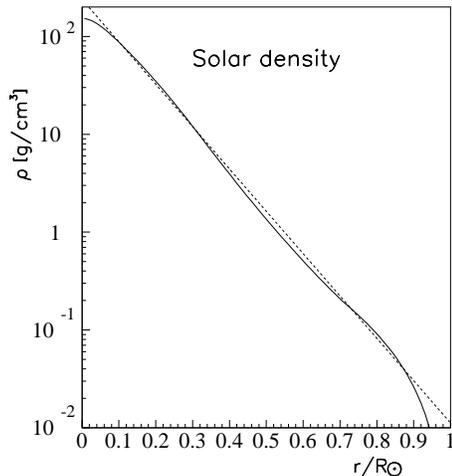}}}
\end{picture}
\vspace*{-12mm}
\caption{Density profile of the sun \protect\cite{BBP-98},
with exponential approximation (dashed).}
\end{center}
\end{figure}
 From the dotted line of Fig.~1, the following values are obtained
\begin{equation}
D(0)=0.0458~{\rm km}^{-1}, \qquad r_0\simeq0.1\times R_\odot.
\end{equation}
The problem is therefore to solve the MSW differential equation,
for $r\ge0$,
\begin{equation}
\label{Eq:Schr-2}
i\,\frac{d}{dr}
\left[\tighten\matrix{
\phi_1(r)\cr \phi_2(r)\cr \phi_3(r)
\cr}\right]
=\frac{1}{2p}
\left[\tighten\matrix{
2pD(0)e^{-r/r_0}+M^2_{11} & M^2_{12} & M^2_{13}\cr
M^2_{21} & M^2_{22} & M^2_{23}\cr
M^2_{31} & M^2_{32} & M^2_{33}
\cr}\right]
\left[\tighten\matrix{
\phi_1(r)\cr \phi_2(r)\cr \phi_3(r)
\cr}\right]
\end{equation}
with the boundary conditions
\begin{equation}
\label{Eq:boundary-cond}
\left[\tighten\matrix{
\phi_1(0)\cr \phi_2(0)\cr \phi_3(0)
\cr}\right]
=
\left[\tighten\matrix{
1\cr 0\cr 0\cr}\right].
\end{equation}

\section{Solution of the Differential Equation}

The differential equation (\ref{Eq:Schr-2}) can be solved explicitly
in terms of generalized hypergeometric functions.
This is to be carried out in the present section.
Indeed, this is the underlying reason for the choice of the exponential
function (\ref{Eq:D-exponential}).

Let
\begin{equation}
u=r/r_0+u_0,
\end{equation}
with
\begin{equation}
\label{Eq:u0}
u_0=-\ln[D(0)r_0]\sim -8.07.
\end{equation}
Then (\ref{Eq:Schr-2}) is
\begin{equation}
\label{Eq:Schr-3}
i\,\frac{d}{du}
\left[\tighten\matrix{
\phi_1\cr \phi_2\cr \phi_3
\cr}\right]
=
\left[\tighten\matrix{
e^{-u}+{\displaystyle \frac{r_0}{2p}}\,M^2_{11} &
{\displaystyle \frac{r_0}{2p}}\,M^2_{12} & {\displaystyle
\frac{r_0}{2p}}\,M^2_{13}\cr \noalign{\vskip4pt}
{\displaystyle
\frac{r_0}{2p}}\,M^2_{21} & {\displaystyle \frac{r_0}{2p}}\,M^2_{22} &
{\displaystyle
\frac{r_0}{2p}}\,M^2_{23}\cr \noalign{\vskip4pt}
{\displaystyle \frac{r_0}{2p}}\,M^2_{31} &
{\displaystyle \frac{r_0}{2p}}\,M^2_{32} & {\displaystyle
\frac{r_0}{2p}}\,M^2_{33}
\cr}\right]
\left[\tighten\matrix{
\phi_1\cr \phi_2\cr \phi_3
\cr}\right].
\end{equation}
Note that $u=0$ corresponds to
\begin{equation}
r=-r_0u_0\sim 0.81 R_\odot.
\end{equation}

In view of the form of Eq.\ (\ref{Eq:Schr-3}), it is convenient to
diagonalize
the $2\times2$ matrix
\begin{displaymath}
\frac{r_0}{2p}
\left[\tighten\matrix{
M^2_{22} & M^2_{23}\cr
M^2_{32} & M^2_{33}
\cr}\right].
\end{displaymath}
Let $\theta_0$ be the angle of rotation that diagonalizes this
$2\times2$ matrix, i.e.,
\begin{equation}
\left[\tighten\matrix{
\cos\theta_0 & -\sin\theta_0\cr
\sin\theta_0 & \cos\theta_0
\cr}\right]
\frac{r_0}{2p}
\left[\tighten\matrix{
M^2_{22} & M^2_{23}\cr
M^2_{32} & M^2_{33}
\cr}\right]
\left[\tighten\matrix{
\cos\theta_0 & \sin\theta_0\cr
-\sin\theta_0 & \cos\theta_0
\cr}\right]
=
\left[\tighten\matrix{
\omega_2 & 0\cr
0 & \omega_3
\cr}\right]
\end{equation}
with
\begin{equation}
\label{Eq:omega-order}
\omega_2 < \omega_3.
\end{equation}
The reason for this Eq.~(\ref{Eq:omega-order}) is that,
if $\omega_2=\omega_3$, then the problem reduces to
neutrino oscillations with only two neutrinos.
Define also
\begin{equation}
\omega_1=\frac{r_0}{2p}\,M^2_{11}, \qquad
\left[\tighten\matrix{
\chi_2\cr \chi_3
\cr}\right]
=\frac{r_0}{2p}
\left[\tighten\matrix{
\cos\theta_0 & -\sin\theta_0\cr
\sin\theta_0 & \cos\theta_0
\cr}\right]
\left[\tighten\matrix{
M^2_{12}\cr M^2_{13}
\cr}\right]
\end{equation}
and
\begin{equation}
\psi_1(u)=\phi_1(u), \qquad
\left[\tighten\matrix{
\psi_2(u)\cr \psi_3(u)
\cr}\right]
=\left[\tighten\matrix{
\cos\theta_0 & -\sin\theta_0\cr
\sin\theta_0 & \cos\theta_0
\cr}\right]
\left[\tighten\matrix{
\phi_2(u)\cr \phi_3(u)
\cr}\right];
\end{equation}
then Eq.~(\ref{Eq:Schr-3}) becomes
\begin{equation}
\label{Eq:Schr-4}
i\,\frac{d}{du}
\left[\tighten\matrix{
\psi_1(u)\cr \psi_2(u)\cr \psi_3(u)
\cr}\right]
=
\left[\tighten\matrix{
\omega_1+e^{-u} & \chi_2 & \chi_3\cr
\chi_2 & \omega_2 & 0\cr
\chi_3 & 0 & \omega_3
\cr}\right]
\left[\tighten\matrix{
\psi_1(u)\cr \psi_2(u)\cr \psi_3(u)
\cr}\right].
\end{equation}

If $\chi_2$ or $\chi_3$ is zero, then once again this reduces
to neutrino oscillations with two generations.
By reversing the signs of $\psi_2(u)$ and/or $\psi_3(u)$ if necessary,
it follows that, without loss of generality
\begin{equation}
\label{Eq:chi-positive}
\chi_j>0, \qquad {\rm for \ \ }j=2,3.
\end{equation}
Let $\mu_1$, $\mu_2$ and $\mu_3$ be the eigenvalues of the $3\times3$
matrix
\begin{displaymath}
\left[\tighten\matrix{
\omega_1 & \chi_2 & \chi_3\cr
\chi_2 & \omega_2 & 0\cr
\chi_3 & 0 & \omega_3
\cr}\right];
\end{displaymath}
then these three $\mu$'s are the squares of the neutrino masses multiplied
by
$r_0/(2p)$. Note that
\begin{equation}
\sum_{j=1}^3 \omega_j
=\sum_{j=1}^3 \mu_j.
\end{equation}
For cases of interest \cite{OW-99-L}, the $\mu_j$ will typically be
large,
$\mu_j\gg 1$. This is of course not a question of units, but rather
a reflection of the physical situation of having a large number
of oscillations within the solar radius.

The interlacing property implies that, by renumbering the $\mu$'s
if necessary,
\begin{equation}
\label{Eq:interlace}
0\le \mu_1 < \omega_2 < \mu_2 < \omega_3 < \mu_3,
\end{equation}
again assuming that there is no reduction to two-neutrino
oscillation.

The following properties are easily verified:
With fixed $\omega_2$, $\omega_3$, $\chi_2$ and $\chi_3$,
\begin{itemize}
\item[(i)]\qquad\vspace{-30pt}
\begin{displaymath}
\frac{\partial\mu_j}{\partial\omega_1}>0 \qquad {\rm for\ \ } j=1,2,3.
\end{displaymath}
\item[(ii)]
As $\omega_1\to\infty$,
\begin{displaymath}
\mu_1\to\omega_2, \qquad \mu_2\to\omega_3, \qquad {\rm and\ \ }
\mu_3\to\infty.
\end{displaymath}
\item[(iii)]
As $\omega_1\to-\infty$, which is physically not possible,
\begin{displaymath}
\mu_1\to-\infty, \qquad \mu_2\to\omega_2, \qquad {\rm and\ \ }
\mu_3\to\omega_3.
\end{displaymath}
\end{itemize}

A third-order ordinary differential equation can be obtained from
Eq.~(\ref{Eq:Schr-4}) as follows.
Define $\psi_{11}$, $\psi_{22}$ and $\psi_{33}$ such that
\begin{eqnarray}
\label{Eq:psi-i}
\psi_1&=&\left(i\,\frac{d}{du}-\omega_2\right)
\left(i\,\frac{d}{du}-\omega_3\right)\psi_{11}, \nonumber \\
\psi_2&=&\chi_2\left(i\,\frac{d}{du}-\omega_3\right)\psi_{22},
\nonumber
\\
\psi_3&=&\chi_3\left(i\,\frac{d}{du}-\omega_2\right)\psi_{33}.
\end{eqnarray}
These $\psi_{11}$, $\psi_{22}$ and $\psi_{33}$ are of course not unique.
It then follows from Eqs.~(\ref{Eq:Schr-4}) and (\ref{Eq:chi-positive})
that
\begin{eqnarray}
\left(i\,\frac{d}{du}-\omega_2\right)
\left(i\,\frac{d}{du}-\omega_3\right)(\psi_{11}-\psi_{22})&=&0,
\nonumber \\
\left(i\,\frac{d}{du}-\omega_2\right)
\left(i\,\frac{d}{du}-\omega_3\right)(\psi_{11}-\psi_{33})&=&0.
\end{eqnarray}
Therefore,
\begin{eqnarray}
\psi_{22}&=&\psi_{11}
+K_{22}e^{-i\omega_2u}+K_{23}e^{-i\omega_3u}, \nonumber \\
\psi_{33}&=&\psi_{11}
+K_{32}e^{-i\omega_2u}+K_{33}e^{-i\omega_3u}.
\end{eqnarray}
In particular,
\begin{equation}
\label{Eq:psi22-psi33}
\psi_{22}-(K_{23}-K_{33})e^{-i\omega_3u}
=\psi_{33}-(K_{32}-K_{22})e^{-i\omega_2u}.
\end{equation}
Define this function of Eq.~(\ref{Eq:psi22-psi33}) to be $\psi$,
then it follows that
\begin{eqnarray}
\psi_{11}&=&\psi-K_{22}e^{-i\omega_2u}-K_{33}e^{-i\omega_3u}, \nonumber \\
\psi_{22}&=&\psi+(K_{23}-K_{33})e^{-i\omega_3u}, \nonumber \\
\psi_{33}&=&\psi+(K_{32}-K_{22})e^{-i\omega_2u}.
\end{eqnarray}
Substitution into Eq.~(\ref{Eq:psi-i}) then expresses $\psi_1$,
$\psi_2$ and $\psi_3$ in terms of the single function $\psi$:
\begin{eqnarray}
\label{Eq:psi1-2-3}
\psi_1&=&\left(i\,\frac{d}{du}-\omega_2\right)
\left(i\,\frac{d}{du}-\omega_3\right)\psi, \nonumber \\
\psi_2&=&\chi_2\left(i\,\frac{d}{du}-\omega_3\right)\psi, \nonumber \\
\psi_3&=&\chi_3\left(i\,\frac{d}{du}-\omega_2\right)\psi.
\end{eqnarray}
Furthermore, this $\psi$ is unique because of (\ref{Eq:omega-order})

To obtain the third-order ordinary differential equation for $\psi$,
it only remains to substitute (\ref{Eq:psi1-2-3}) into the first
equation of (\ref{Eq:Schr-4}):
\begin{eqnarray}
\lefteqn{\left(i\,\frac{d}{du}-\omega_1-e^{-u}\right)
\left(i\,\frac{d}{du}-\omega_2\right)
\left(i\,\frac{d}{du}-\omega_3\right)\psi}\qquad\nonumber \\
&&\mbox{}=\left[\chi_2^2\left(i\,\frac{d}{du}-\omega_3\right)
        +\chi_3^2\left(i\,\frac{d}{du}-\omega_2\right)\right]\psi.
\end{eqnarray}
 From the definition of $\mu_1$, $\mu_2$ and $\mu_3$, this can be
rewritten as
\begin{eqnarray}
\label{Eq:Schr-5}
\lefteqn{\left(i\,\frac{d}{du}-\mu_1\right)
\left(i\,\frac{d}{du}-\mu_2\right)
\left(i\,\frac{d}{du}-\mu_3\right)\psi}\qquad\nonumber \\
&&\mbox{}=e^{-u}
\left(i\,\frac{d}{du}-\omega_2\right)
\left(i\,\frac{d}{du}-\omega_3\right)\psi.
\end{eqnarray}
To put this equation into the form of the differential equation
for the generalized hypergeo\-metric function, let
\begin{equation}
\label{Eq:z-u}
z=ie^{-u};
\end{equation}
then the differential equation (\ref{Eq:Schr-5}) for $\psi$ is
\begin{equation}
\label{Eq:Schr-6}
\left[
\left(z\,\frac{d}{dz}-i\mu_1\right)
\left(z\,\frac{d}{dz}-i\mu_2\right)
\left(z\,\frac{d}{dz}-i\mu_3\right)
-z\left(z\,\frac{d}{dz}-i\omega_2\right)
  \left(z\,\frac{d}{dz}-i\omega_3\right)
\right]\psi=0.
\end{equation}
This is the differential equation for the generalized hypergeometric
function $_2F_2$---see, for example, p.~184 of \cite{Bateman}.
Three linearly independent solutions of this third-order differential
equation (\ref{Eq:Schr-6}) are
\begin{eqnarray}
\psi^{(1)}
&=&K_1e^{-i\mu_1u}
{}_2F_2\left[\left.
\tighten\matrix{
-i(\omega_2-\mu_1), & -i(\omega_3-\mu_1)\cr
1-i(\mu_2-\mu_1),    & 1-i(\mu_3-\mu_1)
\cr}
\right|ie^{-u} \right], \nonumber \\[5pt]
\psi^{(2)}
&=&K_2e^{-i\mu_2u}
{}_2F_2\left[\left.
\tighten\matrix{
-i(\omega_2-\mu_2), & -i(\omega_3-\mu_2)\cr
1-i(\mu_1-\mu_2),    & 1-i(\mu_3-\mu_2)
\cr}
\right|ie^{-u} \right], \nonumber \\[5pt]
\psi^{(3)}
&=&K_3e^{-i\mu_3u}
{}_2F_2\left[\left.
\tighten\matrix{
-i(\omega_2-\mu_3), & -i(\omega_3-\mu_3)\cr
1-i(\mu_1-\mu_3),    & 1-i(\mu_2-\mu_3)
\cr}
\right|ie^{-u} \right],
\end{eqnarray}
where $K_1$, $K_2$ and $K_3$ are arbitrary non-zero constants.
For the convenience of numerical calculations, we make the following choice:
\begin{eqnarray}
K_1&=&\left[(\omega_2-\mu_1)(\omega_3-\mu_1)(\mu_2-\mu_1)(\mu_3-\mu_1)
\right]^{-1/2}, \nonumber \\
K_2&=&\left[(\omega_2-\mu_2)(\omega_3-\mu_2)(\mu_1-\mu_2)(\mu_3-\mu_2)
\right]^{-1/2}, \nonumber \\
K_3&=&\left[(\omega_2-\mu_3)(\omega_3-\mu_3)(\mu_1-\mu_3)(\mu_2-\mu_3)
\right]^{-1/2}.
\end{eqnarray}

The function $_2F_2$ can be defined by the following series
expansion\footnote{From this form, it is also rather obvious
how the confluent hypergeometric function $_1F_1$ emerges as the solution
when one of the $\alpha$'s equals one of the $\rho$'s,
i.e., for the case of two generations.}
\cite{Bateman}
\begin{equation}
\label{Eq:F22-series}
{}_2F_2\left[\left.
\tighten\matrix{
\alpha_1, & \alpha_2\cr
\rho_1,   & \rho_2
\cr}
\right|z \right]
=\sum_{k=0}^\infty
\frac{(\alpha_1)_k(\alpha_2)_k}{(\rho_1)_k(\rho_2)_k}\,
\frac{z^k}{k!},
\end{equation}
where the Pochhammer symbol is defined as
\begin{equation}
(\alpha)_k=\alpha(\alpha+1)\ldots(\alpha+k-1).
\end{equation}

Since the Taylor series of $_2F_2$ about $z=0$, Eq.~(\ref{Eq:F22-series}),
is always convergent,
the differentiations specified in Eqs.~(\ref{Eq:psi1-2-3}) can be carried
out term by term on these Taylor series.
Thus these derivatives of $\psi^{(j)}$ can be easily expressed in
terms of $_2F_2$ with parameters slightly modified.
The result is
\begin{eqnarray}
\label{Eq:psi1-2-3-b}
\psi_1&=&C_1\psi_1^{(1)}+C_2\psi_1^{(2)}+C_3\psi_1^{(3)}, \nonumber \\
\psi_2&=&C_1\psi_2^{(1)}+C_2\psi_2^{(2)}+C_3\psi_2^{(3)}, \nonumber \\
\psi_3&=&C_1\psi_3^{(1)}+C_2\psi_3^{(2)}+C_3\psi_3^{(3)},
\end{eqnarray}
where
\begin{eqnarray}
\label{Eq:psi1}
\psi_1^{(1)}&=&K_1(\mu_1-\omega_2)(\mu_1-\omega_3)e^{-i\mu_1u}
{}_2F_2\left[\left.
\tighten\matrix{
1-i(\omega_2-\mu_1), & 1-i(\omega_3-\mu_1)\cr
1-i(\mu_2-\mu_1),    & 1-i(\mu_3-\mu_1)
\cr}
\right|ie^{-u} \right], \nonumber \\[5pt]
\psi_1^{(2)}&=&K_2(\mu_2-\omega_2)(\mu_2-\omega_3)e^{-i\mu_2u}
{}_2F_2\left[\left.
\tighten\matrix{
1-i(\omega_2-\mu_2), & 1-i(\omega_3-\mu_2)\cr
1-i(\mu_1-\mu_2),    & 1-i(\mu_3-\mu_2)
\cr}
\right|ie^{-u} \right], \nonumber \\[5pt]
\psi_1^{(3)}&=&K_3(\mu_3-\omega_2)(\mu_3-\omega_3)e^{-i\mu_3u}
{}_2F_2\left[\left.
\tighten\matrix{
1-i(\omega_2-\mu_3), & 1-i(\omega_3-\mu_3)\cr
1-i(\mu_1-\mu_3),    & 1-i(\mu_2-\mu_3)
\cr}
\right|ie^{-u} \right],
\end{eqnarray}
\begin{eqnarray}
\label{Eq:psi2}
\psi_2^{(1)}&=&K_1\chi_2(\mu_1-\omega_3)e^{-i\mu_1u}
{}_2F_2\left[\left.
\tighten\matrix{
-i(\omega_2-\mu_1), & 1-i(\omega_3-\mu_1)\cr
1-i(\mu_2-\mu_1),    & 1-i(\mu_3-\mu_1)
\cr}
\right|ie^{-u} \right], \nonumber \\[5pt]
\psi_2^{(2)}&=&K_2\chi_2(\mu_2-\omega_3)e^{-i\mu_2u}
{}_2F_2\left[\left.
\tighten\matrix{
-i(\omega_2-\mu_2), & 1-i(\omega_3-\mu_2)\cr
1-i(\mu_1-\mu_2),    & 1-i(\mu_3-\mu_2)
\cr}
\right|ie^{-u} \right], \nonumber \\[5pt]
\psi_2^{(3)}&=&K_3\chi_2(\mu_3-\omega_3)e^{-i\mu_3u}
{}_2F_2\left[\left.
\tighten\matrix{
-i(\omega_2-\mu_3), & 1-i(\omega_3-\mu_3)\cr
1-i(\mu_1-\mu_3),    & 1-i(\mu_2-\mu_3)
\cr}
\right|ie^{-u} \right],
\end{eqnarray}
\begin{eqnarray}
\label{Eq:psi3}
\psi_3^{(1)}&=&K_1\chi_3(\mu_1-\omega_2)e^{-i\mu_1u}
{}_2F_2\left[\left.
\tighten\matrix{
1-i(\omega_2-\mu_1), & -i(\omega_3-\mu_1)\cr
1-i(\mu_2-\mu_1),    & 1-i(\mu_3-\mu_1)
\cr}
\right|ie^{-u} \right], \nonumber \\[5pt]
\psi_3^{(2)}&=&K_2\chi_3(\mu_2-\omega_2)e^{-i\mu_2u}
{}_2F_2\left[\left.
\tighten\matrix{
1-i(\omega_2-\mu_2), & -i(\omega_3-\mu_2)\cr
1-i(\mu_1-\mu_2),    & 1-i(\mu_3-\mu_2)
\cr}
\right|ie^{-u} \right], \nonumber \\[5pt]
\psi_3^{(3)}&=&K_3\chi_3(\mu_3-\omega_2)e^{-i\mu_3u}
{}_2F_2\left[\left.
\tighten\matrix{
1-i(\omega_2-\mu_3), & -i(\omega_3-\mu_3)\cr
1-i(\mu_1-\mu_3),    & 1-i(\mu_2-\mu_3)
\cr}
\right|ie^{-u} \right].
\end{eqnarray}
In Eqs.~(\ref{Eq:psi1-2-3-b}), the constants $C_1$, $C_2$ and $C_3$
are determined by the boundary conditions Eqs.~(\ref{Eq:boundary-cond}).
For the exponential matter distribution (\ref{Eq:D-exponential}),
this gives completely the solution of Eq.~(\ref{Eq:Schr-1}),
which describes the MSW effect in the sun.

We recall that by Eq.\ (\ref{Eq:u0}), $e^{-u}=r_0D(0)e^{-r/r_0}$.
Thus, the prefactors $e^{-i\mu_j u}$ are proportional to the
exponentials
describing propagation in vacuum, $e^{-i\mu_j r/r_0}$.
In fact, as $r\to\infty$, all these functions approach a simple
limit:
\begin{equation}
\lim_{u\to\infty}{}_2F_2\left[\left.
\tighten\matrix{
\alpha_1, & \alpha_2\cr
\rho_1, & \rho_2
\cr}
\right|ie^{-u} \right]=1.
\end{equation}

While this solution (\ref{Eq:psi1-2-3-b})
is specifically for three generations of neutrinos,
generalization, if desired, to $N$ generations is completely
straightforward, leading to the generalized hypergeometric function
$_{N-1}F_{N-1}$.
Even for three generations, $_2F_2$ is a complicated function.
 From Eq.\ (\ref{Eq:u0}), corresponding to the center of the sun,
$e^{-u}$ is about 3000;
for arguments of this order of magnitude, it is not practical
to calculate the various $_2F_2$ of
Eqs.~(\ref{Eq:psi1})--(\ref{Eq:psi3}) using the Taylor
series expansion.

Sections~IV and~V are devoted to the issue of how these
$_2F_2$ can be calculated.
Of the nine $_2F_2$ that appear in Eqs.~(\ref{Eq:psi1})--(\ref{Eq:psi3}),
three of them are evaluated approximately in Sec.~IV.
In Sec.~V, the remaining six $_2F_2$ are then expressed in terms of
another set of generalized hypergeometric functions $_3F_1$.
These $_3F_1$ have convenient integral representations,
useful for numerical evaluation.
The approach described in the rest of this paper is the best we have found,
but there may well be other methods that are superior.

\section{$_2F_2$ that Appear in $\psi_{\lowercase{j}}^{(3)}$}

In this section, we study the three $_2F_2$ that appear in the
expressions of $\psi_j^{(1)}$, $\psi_j^{(2)}$ and $\psi_j^{(3)}$
as given by Eqs.~(\ref{Eq:psi1})--(\ref{Eq:psi3}). Define
\begin{eqnarray}
\rho_1&=&\mu_3-\mu_1, \qquad \rho_2=\mu_3-\omega_2, \nonumber \\
\rho_3&=&\mu_3-\mu_2, \qquad \rho_4=\mu_3-\omega_3;
\end{eqnarray}
then by (\ref{Eq:interlace})
\begin{equation}
\rho_1 > \rho_2 > \rho_3 > \rho_4 > 0,
\end{equation}
and the three $_2F_2$ of interest are
\begin{eqnarray}
_2F_2(1)&=&
_2F_2\left[\left.
\tighten\matrix{
1+i\rho_2, & 1+i\rho_4\cr
1+i\rho_1, & 1+i\rho_3
\cr}\right|ie^{-u} \right], \nonumber \\[5pt]
_2F_2(2)&=&
_2F_2\left[\left.
\tighten\matrix{
i\rho_2, & 1+i\rho_4\cr
1+i\rho_1, & 1+i\rho_3
\cr}
\right|ie^{-u} \right], \nonumber \\[5pt]
_2F_2(3)&=&
_2F_2\left[\left.
\tighten\matrix{
1+i\rho_2, & i\rho_4\cr
1+i\rho_1, & 1+i\rho_3
\cr}
\right|ie^{-u} \right].
\end{eqnarray}
Let $_2F_2(1)$ be considered in some detail;
the treatments of $_2F_2(2)$ and $_2F_2(3)$ are similar.

 From the definition of $_2F_2$ by its Taylor series,
which is always convergent, it is straightforward to verify, by closing
the contour of integration in the right half-plane, that
\begin{eqnarray}
\label{Eq:F21-contour}
_2F_2(1)&=&\frac{1}{2\pi i}
\,\frac{\Gamma(1+i\rho_1)\Gamma(1+i\rho_3)}{\Gamma(1+i\rho_2)\Gamma(1+i\rho_
4)}
\nonumber \\[5pt]
&&\mbox{}\times\int_{\gamma-i\infty}^{\gamma+i\infty}
\frac{\Gamma(-s)\Gamma(1+i\rho_2+s)\Gamma(1+i\rho_4+s)}
     {\Gamma(1+i\rho_1+s)\Gamma(1+i\rho_3+s)}
\left(e^{-i\pi/2}e^{-u}\right)^s  ds
\end{eqnarray}
where
\begin{equation}
-1 < \gamma < 0.
\end{equation}
This integral is to be evaluated approximately for $u$ large and negative.
Except for an indentation near $s=0$, the contour of integration
can be taken along the imaginary axis.
It is therefore convenient to let
\begin{equation}
s=i\tau.
\end{equation}
The Stirling formula
\begin{equation}
\label{Eq:Stirling-0}
\Gamma(z)\sim(2\pi)^{1/2} e^{-z} e^{(z-1/2)\ln z}
\end{equation}
takes the form
\begin{equation}
\label{Eq:Stirling}
\Gamma(ix)\sim\left(\frac{2\pi}{ix}\right)^{1/2}
e^{-\pi|x|/2} e^{ix(\ln |x|-1)}
\end{equation}
when $z$ is purely imaginary.
Using (\ref{Eq:Stirling}), the integrand in (\ref{Eq:F21-contour}) is
\begin{eqnarray}
\lefteqn{\frac{\Gamma(-s)\Gamma(1+i\rho_2+s)\Gamma(1+i\rho_4+s)}
     {\Gamma(1+i\rho_1+s)\Gamma(1+i\rho_3+s)}
\left(e^{-i\pi/2}e^{-u}\right)^s}\qquad\nonumber \\[5pt]
&=&\frac{\Gamma(-i\tau)\Gamma(1+i(\tau+\rho_2))\Gamma(1+i(\tau+\rho_4))}
     {\Gamma(1+i(\tau+\rho_1))\Gamma(1+i(\tau+\rho_3))}\,
e^{\pi\tau/2}e^{i|u|\tau} \nonumber \\[5pt]
&\sim&\left(\frac{2\pi}{-i\tau}\right)^{1/2}
\frac{[i(\tau+\rho_2)]^{1/2}[i(\tau+\rho_4)]^{1/2}}
     {[i(\tau+\rho_1)]^{1/2}[i(\tau+\rho_3)]^{1/2}}
\, e^{\pi\theta_1(\tau)/2}\,e^{i\theta_2(\tau)},
\end{eqnarray}
with
\begin{equation}
\theta_1(\tau)=|\tau+\rho_1|-|\tau+\rho_2|+|\tau+\rho_3|-|\tau+\rho_4|
-|\tau|+\tau,
\end{equation}
and
\begin{eqnarray}
\theta_2(\tau)
&=&-\tau(\ln|\tau|-1)-(\tau+\rho_1)(\ln|\tau+\rho_1|-1)
+(\tau+\rho_2)(\ln|\tau+\rho_2|-1) \nonumber \\
& &-(\tau+\rho_3)(\ln|\tau+\rho_3|-1)
+(\tau+\rho_4)(\ln|\tau+\rho_4|-1)+|u|\tau.
\end{eqnarray}
The condition
\begin{equation}
\frac{\partial\theta_2(\tau)}{\partial\tau}=0
\end{equation}
for stationary phase gives
\begin{equation}
\label{Eq:stat-phase}
\frac{|\tau||\tau+\rho_1||\tau+\rho_3|}{|\tau+\rho_2||\tau+\rho_4|}
=e^{|u|}.
\end{equation}
Eq.~(\ref{Eq:stat-phase}) together with the fact that $\theta_1(\tau)$
is a non-decreasing function of $\tau$ implies that the relevant point
of stationary phase occurs when $\tau$ is positive.
Let $\tau_0$ be the unique positive root of the cubic equation
\begin{equation}
\frac{\tau_0(\tau_0+\rho_1)(\tau_0+\rho_3)}{(\tau_0+\rho_2)(\tau_0+\rho_4)}
=e^{|u|},
\end{equation}
then $\tau_0$ is the point of stationary phase that gives the leading
contribution to $_2F_2(1)$. The result is, with all gamma functions
replaced by the corresponding Stirling formula (\ref{Eq:Stirling-0}),
\begin{eqnarray}
\label{Eq:F22-Stirling-1}
_2F_2(1)
&\sim&\left(\frac{\rho_1\rho_3}{\rho_2\rho_4}\right)^{1/2}
\left[\frac{(\tau_0+\rho_2)(\tau_0+\rho_4)}
           {(\tau_0+\rho_1)(\tau_0+\rho_3)}\right]^{1/2}\,
e^{i[\rho_1(\ln\rho_1-1)-\rho_2(\ln\rho_2-1)
    +\rho_3(\ln\rho_3-1)-\rho_4(\ln\rho_4-1)]}
\nonumber \\
&&\mbox{}\times
\left[1+\frac{\tau_0}{\tau_0+\rho_1}-\frac{\tau_0}{\tau_0+\rho_2}
+\frac{\tau_0}{\tau_0+\rho_3}-\frac{\tau_0}{\tau_0+\rho_4}\right]^{-1/2}
e^{i\theta_2(\tau_0)},
\end{eqnarray}
for $u$ negative and large.
Similarly,
\begin{eqnarray}
\label{Eq:F22-Stirling-2}
_2F_2(2)
&\sim&\frac{\rho_2}{\tau_0+\rho_2}\, _2F_2(1), \\[5pt]
\label{Eq:F22-Stirling-3}
_2F_2(3)
&\sim&\frac{\rho_4}{\tau_0+\rho_4}\, _2F_2(1).
\end{eqnarray}
These are the desired results.

\section{$_2F_2$ that Appear in $\psi_{\lowercase{j}}^{(1)}$ and
$\psi_{\lowercase{j}}^{(2)}$}

The next step is to study the $_2F_2$ that appear in the expressions
(\ref{Eq:psi1})--(\ref{Eq:psi3}) for $\psi_1^{(i)}$, $\psi_2^{(i)}$
and $\psi_3^{(i)}$ with $i=1,2$.
It should be emphasized that, while the treatment in the preceding
Sec.~IV involves approximations, what is to be carried out in this section
is exact. Since exact manipulations are usually more straightforward,
the description will be relatively more brief in this section.

Consider the third-order ordinary differential equation
\begin{equation}
\label{Eq:Schr-7}
\left[
\left(z\,\frac{d}{dz}+\beta_1\right)
\left(z\,\frac{d}{dz}+\beta_2\right)
\left(z\,\frac{d}{dz}+\beta_3\right)
-z\left(z\,\frac{d}{dz}+\alpha_1\right)
  \left(z\,\frac{d}{dz}+\alpha_2\right)
\right]f=0.
\end{equation}
A comparison with Eq.~(\ref{Eq:Schr-6}) shows that
\begin{equation}
\label{Eq:beta-mu}
\beta_1=-i\mu_1, \qquad
\beta_2=-i\mu_2, \qquad
\beta_3=-i\mu_3.
\end{equation}
The values of $\alpha_1$ and $\alpha_2$ are different for the various
$\psi$'s:
\begin{eqnarray}
\label{Eq:psi-1-alphas}
\text{for}\
\psi_1^{(j)}\text{:}\qquad\alpha_1=1-i\omega_2&\quad\text{and}\quad
\alpha_2=1-i\omega_3,\\
\text{for}\
\psi_2^{(j)}\text{:}\qquad\alpha_1=-i\omega_2\hspace{11pt}&\quad
\text{and}\quad \alpha_2=1-i\omega_3,\\
\text{for}\
\psi_3^{(j)}\text{:}\qquad \alpha_1=1-i\omega_2&\quad
\text{and}\quad \alpha_2=-i\omega_3,\hspace{11pt}\\
\text{for}\
\psi^{(j)}\text{:}\qquad\alpha_1=-i\omega_2\hspace{11pt}&\quad
\text{and}\quad \alpha_2=-i\omega_3.\hspace{11pt}\end{eqnarray}
The three linearly independent solutions of Eq.~(\ref{Eq:Schr-7}) are
\begin{eqnarray}
\label{Eq:f1-f2-f3}
f^{(1)}(z)&=&z^{-\beta_1}
{}_2F_2\left[\left.
\tighten\matrix{
\alpha_1-\beta_1, & \alpha_2-\beta_1\cr
1+\beta_2-\beta_1, & 1+\beta_3-\beta_1
\cr}
\right|z \right], \nonumber \\[5pt]
f^{(2)}(z)&=&z^{-\beta_2}
{}_2F_2\left[\left.
\tighten\matrix{
\alpha_1-\beta_2, & \alpha_2-\beta_2\cr
1+\beta_1-\beta_2, & 1+\beta_3-\beta_2
\cr}
\right|z \right], \nonumber \\[5pt]
f^{(3)}(z)&=&z^{-\beta_3}
{}_2F_2\left[\left.
\tighten\matrix{
\alpha_1-\beta_3, & \alpha_2-\beta_3\cr
1+\beta_1-\beta_3, & 1+\beta_2-\beta_3
\cr}
\right|z \right],
\end{eqnarray}
where $f^{(3)}(z)$ is the function treated in the preceding Sec.~IV.

On the other hand, if in Eq.~(\ref{Eq:Schr-7}) the independent variable
is changed to
\begin{equation}
\hat z = z^{-1},
\end{equation}
then the differential equation is
\begin{equation}
\label{Eq:Schr-8}
\left[
\left(\hat z\,\frac{d}{d\hat z}-\alpha_1\right)
\left(\hat z\,\frac{d}{d\hat z}-\alpha_2\right)
+\hat z\left(\hat z\,\frac{d}{d\hat z}-\beta_1\right)
       \left(\hat z\,\frac{d}{d\hat z}-\beta_2\right)
       \left(\hat z\,\frac{d}{d\hat z}-\beta_3\right)
\right]f=0.
\end{equation}
In this form, two of the three linearly independent solutions of
Eq.~(\ref{Eq:Schr-7}) are (in terms of $z$):
\begin{eqnarray}
\label{Eq:hat-f}
\hat f^{(1)}(z)
&=&z^{-\alpha_1}
{}_3F_1\left[\left.
\tighten\matrix{
-\beta_1+\alpha_1, & -\beta_2+\alpha_1, & -\beta_3+\alpha_1\cr
1-\alpha_2+\alpha_1 & &
\cr}
\right|-z^{-1} \right], \nonumber \\[5pt]
\hat f^{(2)}(z)
&=&z^{-\alpha_2}
{}_3F_1\left[\left.
\tighten\matrix{
-\beta_1+\alpha_2, & -\beta_2+\alpha_2, & -\beta_3+\alpha_2\cr
1-\alpha_1+\alpha_2 & &
\cr}
\right|-z^{-1} \right].
\end{eqnarray}
For a discussion of the generalized hypergeometric functions
$_pF_q$ with $p>q+1$, see for example Chapter~V of \cite{Bateman}.

It should be added parenthetically that, for the present purposes,
$z^{-\beta_1}$, as an example, can sometimes be very small because
by Eqs.~(\ref{Eq:z-u}) and (\ref{Eq:beta-mu})
\begin{equation}
z^{-\beta_1}=e^{-\pi\mu_1/2}\,e^{-i\mu_1u}.
\end{equation}
Thus, when $\mu_1$ is large, which is often the case,
for numerical calculations it is convenient to replace
$z^{-\beta_1}$ by $e^{-i\mu_1u}$, and similarly for the other powers
of $z$ multiplying the $_2F_2$ and $_3F_1$.

Since Eq.~(\ref{Eq:Schr-7}) is a third-order linear ordinary
differential equation, the two solutions as given by Eq.~(\ref{Eq:hat-f})
must be expressible as linear combinations of the three solutions of
Eq.~(\ref{Eq:f1-f2-f3}). The result is
\begin{eqnarray}
\label{Eq:hat-f1}
\hat f^{(1)}(z)
&=&\frac{\pi\Gamma(1-\alpha_2+\alpha_1)}
{\Gamma(-\beta_1+\alpha_1)\Gamma(-\beta_2+\alpha_1)\Gamma(-\beta_3+\alpha_1)
}
\nonumber \\[4pt]
&&\mbox{}\times\left[
 \frac{\Gamma(-\beta_2+\beta_1)\Gamma(-\beta_3+\beta_1)}
{\Gamma(1-\alpha_1+\beta_1)\Gamma(1-\alpha_2+\beta_1)}
\,\frac{1}{\sin\pi(\alpha_1-\beta_1)}\, f^{(1)}(z)\right. \nonumber \\[4pt]
&&\mbox{}+\frac{\Gamma(-\beta_1+\beta_2)\Gamma(-\beta_3+\beta_2)}
{\Gamma(1-\alpha_1+\beta_2)\Gamma(1-\alpha_2+\beta_2)}
\,\frac{1}{\sin\pi(\alpha_1-\beta_2)}\, f^{(2)}(z) \nonumber \\[4pt]
&&\mbox{}+\left.\frac{\Gamma(-\beta_1+\beta_3)\Gamma(-\beta_2+\beta_3)}
{\Gamma(1-\alpha_1+\beta_3)\Gamma(1-\alpha_2+\beta_3)}
\,\frac{1}{\sin\pi(\alpha_1-\beta_3)}\, f^{(3)}(z) \right],
\end{eqnarray}
\begin{equation}
\label{Eq:hat-f2}
\hat f^{(2)}(z)=\hat f^{(1)}(z)\Big|_{\alpha_1\leftrightarrow\alpha_2}.
\end{equation}

Eqs.~(\ref{Eq:hat-f1}) and (\ref{Eq:hat-f2}) can be considered
to be the formulas that express $f^{(1)}(z)$ and $f^{(2)}(z)$
in terms of $\hat f^{(1)}(z)$,  $\hat f^{(2)}(z)$ and $f^{(3)}(z)$.
Since $f^{(3)}(z)$ is given
by Eqs.~(\ref{Eq:F22-Stirling-1})--(\ref{Eq:F22-Stirling-3}), it only
remains to obtain $\hat f^{(1)}(z)$ and $\hat f^{(2)}(z)$,
which are given by $_3F_1$ rather than $_2F_2$.

The advantage of $_3F_1$ over $_2F_2$ is due to the integral
representation
\begin{equation}
\label{Eq:F31-int-rep-1}
_3F_1\left[\left.
\tighten\matrix{
a_1, & a_2, & a_3 \cr
b &  &
\cr}
\right|-x^{-1} \right]
=\frac{1}{\Gamma(a_1)}\,x^{a_1}
\int_0^\infty dt\, e^{-xt}\, t^{a_1-1}\,
{}_2F_1(a_2,a_3;b;-t),
\end{equation}
where $_2F_1$ is the usual hypergeometric function.
This Eq.~(\ref{Eq:F31-int-rep-1}) follows from Eq.~(8) on p.~214
and Eq.~(1) on p.~215 of \cite{Bateman},
and has the great advantage of not containing in the integrand
any gamma function that depends on the variable $t$,
because such gamma functions require excessive computer time.

For the $_2F_1$ in the integrand of (\ref{Eq:F31-int-rep-1}),
the representation
\begin{eqnarray}
\label{Eq:F21-Pochhammer}
_2F_1(a,b;c;-t)
&=&\frac{-\Gamma(c)e^{-i\pi c}}
        {4\Gamma(b)\Gamma(c-b)\sin\pi b\sin\pi(c-b)} \nonumber \\[5pt]
&&\mbox{}\times\int_{\cal P} s^{b-1}(1-s)^{c-b-1}(1+ts)^{-a}ds
\end{eqnarray}
is convenient, where ${\cal P}$ is the Pochhammer contour of Fig.~2.
In the numerical evaluation through Eqs.~(\ref{Eq:F31-int-rep-1})
and (\ref{Eq:F21-Pochhammer}), it is convenient to deform
the contours of integration through the point of stationary phase.
This point is discussed in more detail in the Appendix.
\begin{figure}[htb]
\refstepcounter{figure}
\label{Fig:Pochhammer}
\addtocounter{figure}{-1}
\begin{center}
\setlength{\unitlength}{1cm}
\begin{picture}(8,7.0)
\put(-3.,2.0)
{\mbox{\epsfysize=4.0cm\epsffile{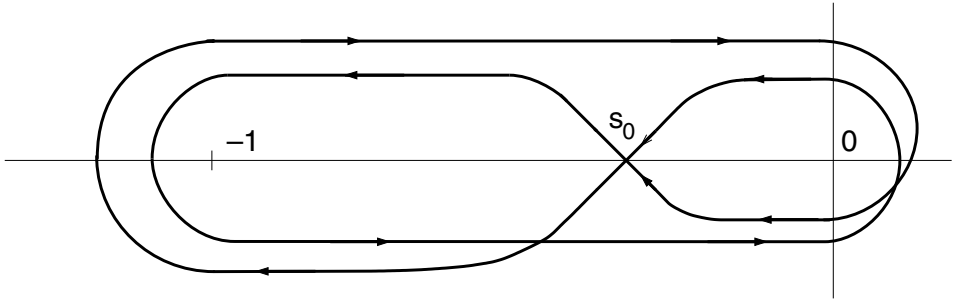}}}
\end{picture}
\vspace*{-8mm}
\caption{Pochhammer contour used for the evaluation of
the hypergeometric function $_2F_1$.
The point of stationary phase, denoted $s_0$, occurs along the path from NE
to SW.}
\end{center}
\end{figure}

\section{Summary and Discussion}

The Mikheyev-Smirnov-Wolfenstein effect in the sun has been studied and
the differential equations solved
for three types of neutrinos coupled through their mass matrix.
Under the assumption, as expressed by Eq.~(\ref{Eq:D-exponential}),
that the electron density can be approximated by an exponential function,
the MSW differential equations are solved exactly in Sec.~III, especially
Eqs.~(\ref{Eq:psi1-2-3-b})--(\ref{Eq:psi3}), in terms of the generalized
hypergeometric function $_2F_2$.
The method used there can be immediately generalized to $N$ types
of neutrinos coupled in the same way, leading to $_{N-1}F_{N-1}$.

Since $_2F_2$ cannot be considered to be a familiar function,
Secs.~IV and~V are devoted to the issue of how they can be calculated.
Of the nine $_2F_2$ that appear in Eqs.~(\ref{Eq:psi1})--(\ref{Eq:psi3}),
three are evaluated approximately in Sec.~IV.
In Sec.~V, the remaining six $_2F_2$ are then expressed in terms of
another set of generalized hypergeometric functions, $_3F_1$.
These $_3F_1$ have convenient integral representations given
by Eqs.~(\ref{Eq:F31-int-rep-1}) and (\ref{Eq:F21-Pochhammer}),
useful for numerical evaluation.
Eqs.~(\ref{Eq:F31-int-rep-1}) and (\ref{Eq:F21-Pochhammer}) can also
be used to get asymptotic expressions for $_3F_1$ and hence $_2F_2$,
but this development is not discussed here because it is not needed
for the study of neutrino oscillations \cite{OW-99-L}.

Finally, it should be mentioned that, while the calculation of Sec.~IV
can be generalized immediately to $N$ types of neutrinos,
there does not seem to be a similar straightforward generalization
for Sec.~V.
More precisely, while $_{N-1}F_{N-1}$ can be related to $_{N}F_{N-2}$,
the generalization of Eq.~(\ref{Eq:F31-int-rep-1}) involves
$_{N-1}F_{N-2}$, which does not have an integral representation
similar to Eq.~(\ref{Eq:F21-Pochhammer}) involving a single integral.
We believe that, for $_{N}F_{N-2}$, it is necessary to use an $(N-1)$-fold
integral to avoid having gamma functions that depend on the variables
of integration.

\acknowledgments{This paper owes its existence to the late 
Professor Harry Lehmann. 
He introduced us to the issues of mass matrices 
and participated in the early stages of this work.
Also, we would like to thank Professors Evgeny Akhmedov, Alvaro de Rujula, 
Alexei Smirnov and Gabriele Veneziano, and especially
Conrad Newton, for useful discussions.
T. T. Wu wishes to thank the Theory Division of CERN for its kind
hospitality.
\hfil\break
\indent This work was supported in part by the Research
Council of Norway, and in part by the United
States Department of Energy under Grant No.\ DE-FG02-84ER40158.}

\eject
\appendix

\section*{}

In this Appendix, Eqs.~(\ref{Eq:F31-int-rep-1}) and
(\ref{Eq:F21-Pochhammer})
are studied further. For definiteness, let $\alpha_1$ and $\alpha_2$
be defined by (\ref{Eq:psi-1-alphas}), and consider the $_3F_1$
that appears in the first equation in (\ref{Eq:hat-f}).
Use the inequality (\ref{Eq:interlace}) to define the four positive
quantities
\begin{eqnarray}
\label{Eq:xi-eta-zeta-zetap}
\xi   &=&\mu_2-\omega_2, \nonumber \\
\eta  &=&\omega_2-\mu_1, \nonumber \\
\zeta &=&\mu_3-\omega_2, \nonumber \\
\zeta'&=&\omega_3-\omega_2 < \zeta;
\end{eqnarray}
then Eq.~(\ref{Eq:F31-int-rep-1}) gives, for this $_3F_1$ under
consideration,
\begin{eqnarray}
\label{Eq:F31-int-rep-2}
\lefteqn{_3F_1\left[\left.
\tighten\matrix{
1+i\xi, & 1-i\eta, & 1+i\zeta \cr
1+i\zeta', &  &
\cr}
\right|\frac{i}{y} \right] }\qquad\nonumber \\[5pt]
&&\mbox{}=\frac{1}{\Gamma(1+i\xi)}\,\left(e^{i\pi/2}y\right)^{1+i\xi}
\int_0^\infty dt\, e^{-ity}\, t^{i\xi}\,
{}_2F_1(1-i\eta,1+i\zeta;1+i\zeta';-t),
\end{eqnarray}
where, by Eq.~(\ref{Eq:z-u}),
\begin{equation}
y=e^{-u}
\end{equation}
is large for $u$ negative and large. 
The substitution of Eq.~(\ref{Eq:F21-Pochhammer}) into
Eq.~(\ref{Eq:F31-int-rep-2}) then gives
\begin{eqnarray}
\label{Eq:F31-int-rep-3}
\lefteqn{_3F_1\left[\left.
\tighten\matrix{
1+i\xi, & 1-i\eta, & 1+i\zeta \cr
1+i\zeta', &  &
\cr}
\right|\frac{i}{y} \right]}\qquad \nonumber \\[5pt]
&&\mbox{}=\frac{1}{\Gamma(1+i\xi)}\,\left(e^{i\pi/2}y\right)^{1+i\xi}
\frac{\Gamma(1+i\zeta')e^{-i\pi(1+i\zeta')}}
{\Gamma(1-i\eta)\Gamma(i(\eta+\zeta'))4\sin\pi(1-i\eta)\sin\pi
i(\eta+\zeta')}
\, I,
\end{eqnarray}
where
\begin{eqnarray}
\label{Eq:I-t-s}
I&=&\int_0^\infty dt \int_{\cal P} ds\, e^{-ity}\, t^{i\xi}\,
s^{-i\eta}(1+s)^{i(\zeta-\zeta')}(1+s+st)^{-1-i\zeta} \nonumber \\[5pt]
&=&\int_0^\infty dt \int_{\cal P} ds\, (1+s+st)^{-1}\,
e^{-i\phi(t,s)},
\end{eqnarray}
with
\begin{equation}
\phi(t,s)=ty-\xi\ln t +\eta\ln s-(\zeta-\zeta')\ln(1+s)+\zeta\ln(1+s+st).
\end{equation}
This $\phi$ is of course not related to the $\phi_j$ of
Eq.~(\ref{Eq:Schr-1}).

For the double integral in Eq.~(\ref{Eq:I-t-s}), the point or points of
stationary phase are given by
\begin{equation}
\frac{\partial\phi(t,s)}{\partial t}
=\frac{\partial\phi(t,s)}{\partial s}=0,
\end{equation}
or
\begin{equation}
\label{Eq:stat-cond-1}
y-\frac{\xi}{t}+\frac{s\zeta}{1+s+st}=0,
\end{equation}
and
\begin{equation}
\label{Eq:stat-cond-2}
\frac{\eta}{s}-\frac{\zeta-\zeta'}{1+s}+\frac{(1+t)\zeta}{1+s+st}=0.
\end{equation}
Elimination of $s$ between Eqs.~(\ref{Eq:stat-cond-1})
and (\ref{Eq:stat-cond-2}) gives the cubic equation for $t$,
\begin{equation}
\label{Eq:ell-cubic}
\ell(t)=0,
\end{equation}
with
\begin{equation}
\ell(t)=y^2t^3+y(y+\zeta-2\xi-\eta)t^2
+[(\zeta'-2\xi)y-(\zeta-\xi)(\xi+\eta)]t
-\xi(\zeta'-\xi).
\end{equation}
When $\zeta'\to\zeta$, the three roots of Eq.~(\ref{Eq:ell-cubic})
are explicitly
\begin{eqnarray}
t&=&t_+=(2y)^{-1}\bigl\{-(y-\xi-\eta)+[(y-\xi-\eta)^2+4\xi y]^{1/2}
\bigr\} > 0, \\
t&=&t_-=(2y)^{-1}\bigl\{-(y-\xi-\eta)-[(y-\xi-\eta)^2+4\xi y]^{1/2}
\bigr\} < 0, \\
t&=&-(\zeta-\xi)/y < 0.
\end{eqnarray}
Since by (\ref{Eq:xi-eta-zeta-zetap})
\begin{equation}
\ell(t_-)=-\frac{\eta t_-}{1+t_-}\,(\zeta-\zeta') > 0,
\end{equation}
we have
\begin{equation}
\label{Eq:t0-cases}
\ell(t)\tighten\cases{
< 0, & for $t$ sufficiently large and negative,\cr
> 0, & for $t=t_-<0$,\cr
< 0, & for $t=0$,\cr
> 0, & for $t$ sufficiently large and positive.\cr
}
\end{equation}
It therefore follows that the cubic equation (\ref{Eq:ell-cubic})
has one and only one positive root.
Since the $t$ integral in Eq.~(\ref{Eq:I-t-s}) is along the positive
real axis, this implies that there is exactly one relevant point
of stationary phase, say ($t_0,s_0$), with $t_0$ given by this unique
positive root of Eq.~(\ref{Eq:ell-cubic}).
Furthermore, since
\begin{equation}
\ell(t_+) < 0,
\end{equation}
it follows from (\ref{Eq:t0-cases}) that
\begin{equation}
t_0 > t_+.
\end{equation}

 From Eq.~(\ref{Eq:stat-cond-1}), the corresponding $s_0$ is given by
\begin{equation}
s_0 =-\left[1+t_0+\frac{\zeta t_0}{yt_0-\xi}\right]^{-1},
\end{equation}
and hence
\begin{equation}
-(1+t_0)^{-1} < s_0 < 0.
\end{equation}
For the numerical evaluation of this $_3F_1$ of
Eq.~(\ref{Eq:F31-int-rep-3}),
using Eq.~(\ref{Eq:I-t-s}), it is convenient and efficient to choose
the Pochhammer contour ${\cal P}$ to go through this value of $s_0$,
and also to deform the contour for the $t$ integration so that at $t_0$
it is locally the path of steepest descent.

The other $_3F_1$'s can be treated in similar ways.

\end{document}